\documentclass[
nobibnotes,aps,pre,showkeys,
12pt]{revtex4-2}
\usepackage{hyperref}
\usepackage{graphics}
\usepackage{subcaption}
\usepackage{graphicx}
\usepackage{color}
\usepackage{amsmath}
\usepackage{amssymb}
\usepackage{bm}
\usepackage{float}
\usepackage{epsfig}
\usepackage{epstopdf}
\usepackage{color,graphics}
\usepackage{mathtools}

\begin{document}

\title{Compton Scattering of Plasmons}

\author{J. Tito Mendon\c{c}a}

\email{titomend@tecnico.ulisboa.pt}

\affiliation{GoLP / IPFN, Instituto Superior T\'{e}cnico, Universidade de Lisboa, Av. Rovisco Pais 1, 1049-001 Lisboa, Portugal.}

\author{Fernando Haas}

\email{fernando.haas@ufrgs.br}

\affiliation{Physics Institute, Federal University of Rio Grande do Sul, Av. Bento Gon\c calves 9500, 91501-970 Porto Alegre RS, Brazil.}

\begin{abstract}

We extend de concept of Compton scattering to the case of plasmons. This concept was originally applied to electrons in vacuum. Here, we consider  electrons in a plasma, and study the scattering properties of photon-plasmon interactions. We show that a number $n$ of plasmons with frequency $ \omega \simeq \omega_p$ is scattered by an electron, for an incident photon with frequency $\omega' \geq \omega_p$, where $\omega_p$ is the plasma frequency. We describe the general case of arbitrary $n$  and assume that Compton scattering of plasmons is intrinsically a nonlinear process. Our theoretical model is based on Volkov solutions of the Klein-Gordon equation describing the state of relativistic electrons, when the spin is ignored. We derive the corresponding scattering probability, as well as the recoil formula associated with arbitrary the final electron states. This process can be relevant to intense laser plasma interactions.

\end{abstract}


\maketitle

\section{Introduction}

Compton scattering is a basic elementary process of radiation interactions with matter. First discovered in 1923, using a theoretical interpretation of X-ray experiments \cite{compton}, it played an important historical role in the understanding of the photon concept and in the establishment of the modern quantum theory of light \cite{heitler,itzykson,mbook}. 
In its original version, it explains the increase of wavelength upon scattering by an electron, as stated by the famous Compton recoil formula. But, the inverse process, where the photon energy increases and emission of high frequency photons results from scattering of low energy ones (such as those associated with the cosmic microwave background) on energetic electrons is also possible \cite{blumenthal}. This inverse process can be very important for the understanding of high-energy astrophysics \cite{zhang,hartemann}.  A further extension of the Compton scattering concept is the nonlinear scattering regime  \cite{brown,nikishov,eberly}, which is relevant to intense plasma interactions in the very high intensity limit \cite{dipiazza,seipt,thoma}. Strong evidence of nonlinear Compton scattering is expected with the new Peta-Watt laser systems \cite{corels}.

Here, we extend de concept of Compton scattering to the case of plasmons. It is well-known that this concept was originally applied to photons interacting with electrons in vacuum. Here, we consider the interaction of photons with electrons in a plasma, and study the scattering properties of this interaction. We show that a number of plasmons is scattered by an electron, for an incident photon with frequency $\omega'$. For thermal electrons, this number is of the order of the ratio between the photon frequency and the plasma frequency, $n \simeq \omega' / \omega_p$. For this reason, Compton scattering of photons is intrinsically a nonlinear process. 

Our theoretical model is based on the use of Volkov solutions \cite{volkov}. Originally,  these solutions were derived for electrons in the presence of an electromagnetic wave in vacuum. In recent years, the Volkov solutions were extended to the case of electrons in a plasma \cite{pre2011,eliezer,haas} where, not only the dispersion properties of the electromagnetic waves have to be taken into account, but more importantly, electrostatic waves can also be considered. Here we use the Klein-Gordon equation describing the behaviour of relativistic electron states, when the spin is ignored. We derive the corresponding scattering probability and the recoil formula describing the final electron state, associated with Compton scattering with emission of $n \geq 1$ plasmons. This process can be relevant to intense laser plasma interactions.

\section{Volkov Solutions}

We start with the Klein-Gordon equation describing the wavefunction $\psi$ of a spinless electron in the presence of two waves, one electromagnetic (representing a laser pulse) and the other electrostatic (representing an electron plasma oscillation). We use
\begin{equation}
\left[ \left( i \partial^\mu + k_C a^\mu \right)^2 - k_C^2 \right] \psi = 0 \, ,
\label{2.1} \end{equation}
where $a^\mu \equiv ( U, {\bf a} )$, and $\partial^\mu \equiv ( \partial_t / c, - \nabla)$. Here, $k_C = m c / \hbar$ is the Compton wavenumber and the normalized scalar and vector potentials are 
\begin{equation}
U = \frac{e V}{m c^2} \, , \quad {\bf a} = \frac{e {\bf A}}{m c} \, .
\label{2.1b} \end{equation}
where $V$ and ${\bf A}$ are the usual electromagnetic potentials, and $-e$ and $m$ the electron charge and mass. 
We use the metric signature $(+, -, -, -)$ and adopt the Lorentz gauge. Therefore 
\begin{equation}
a_\mu a^\mu = U^2 - {\bf a}^2 \, , \quad \partial_\mu a^\mu = \frac{1}{c} \partial_t U + \nabla \cdot {\bf a} = 0 \, .
\label{2.1c} \end{equation}
In more explicit form, we can write eq. (\ref{2.1}) as
\begin{eqnarray}
\left[ \left( \nabla^2 - \frac{1}{c^2} \partial_t^2 \right) + 2 \frac{i}{c} k_C \left( U \partial_t + c \, {\bf a} \cdot \nabla \right) - \right. \nonumber \\
 - \left. \left(1 + {\bf a}^2 - U^2 \right) k_C^2 \right] \psi = 0 \, .
\label{2.2} \end{eqnarray}
This is the starting point of our model, where spin effects are ignored. In order to account for spin, we would need to the replace the above Klein-Gordon equation (\ref{2.1})  by the quadratic Dirac equation, which takes the form 
\begin{equation}
\left[ \left( i \partial^\mu + k_C a^\mu \right)^2 + k_C^2 \left(i \bar{\bar{\alpha}} \cdot {\bf e} + \bar{\bar{\Sigma}} \cdot {\bf b} \right) - k_C^2 \right] \psi = 0 \, ,
\label{2.2b} \end{equation}
where $\psi$ is now a 4-spinor, $\bar{\bar{\alpha}}$ is the usual  Dirac matrix, and $\bar{\bar{\Sigma}} = - (i/2) (\bar{\bar{\alpha}} \times \bar{\bar{\alpha}})$. Here, we notice the appearance of new terms, containing the normalized electric and magnetic fields ${\bf e}= - \partial_t {\bf a} / c - \nabla U$ and ${\bf b} = \nabla \times {\bf a}$. These new terms can describe spin coupling as well as electron-positron effects. A similar, although formally more complicated, calculation could be done with this equation. Here, we consider the case of isotropic plasmas and moderately high laser field configurations, where these affects can usually be ignored and the use of eq. (\ref{2.1}) is justified.

We consider two waves, an electrostatic wave with frequency and wavevector $(\omega, {\bf k})$, and an electromagnetic wave $(\omega', {\bf k'})$, such that $\omega' > \omega \simeq \omega_p$, where $\omega_p$ is the electron plasma frequency. These waves are described by the 4-potential $a^\mu \equiv \left[ U (\tau), {\bf a} (\tau') \right]$, such that $U ({\bf r}, t) = U_0 f (\tau)$ and ${\bf a} ({\bf r}, t) = {\bf a}_0 f' (\tau')$, using the time variables $\tau = t - ({\bf k} \cdot {\bf r}) / \omega$ and $\tau' = t - ({\bf k'} \cdot {\bf r}) / \omega'$. Here, $U_0$ and ${\bf a}_0$ are constant amplitudes, and $f (\tau)$ and $f' (\tau')$ are arbitrary oscillating functions to be specified. 
Following the standard Volkov approach \cite{volkov,itzykson}, we solve eq. (\ref{2.2}) using a solution of the form
\begin{equation}
\psi ({\bf r}, t) = e^{i \theta} \Phi (\tau, \tau') \, , 
\label{2.3} \end{equation}
where
\begin{equation}
\theta = - \frac{1}{\hbar} p^\mu x_\mu = - \frac{1}{\hbar} \left( \epsilon_e t - {\bf p}_e \cdot {\bf r} \right) \, .
\label{2.3b} \end{equation}
Here, we have used the 4-momentum $p^\mu = (\epsilon_e /c , {\bf p}_e)$, where $\epsilon_e$ and ${\bf p}_e$ are the electron energy and momentum. Replacing this in eq. (\ref{2.2}), and using the effective mass-gap equation 
\begin{equation}
\epsilon_e / c = \sqrt{p_e^2 + m^2 c^2 (1 + \left< a^2 \right> )} \, ,
\label{2.3c} \end{equation}
where $< a^2 >$ represents the time-average over one cycle $2 \pi / \omega'$, after following a straightforward calculation, we arrive at an evolution equation for the reduced wavefunction $\Phi (\tau, \tau')$, of the form
\begin{equation}
2 i g (\tau, \tau') \frac{\partial \Phi}{\partial \tau} + 2 i g' (\tau, \tau') \frac{\partial \Phi}{\partial \tau'} + F (\tau, \tau') \Phi = 0 \, .
\label{2.4} \end{equation}
In this equation we have introduced new functions, defined by
\begin{equation}
g (\tau, \tau') = \omega_e + c k_C \left( U - \frac{c}{\omega} {\bf a} \cdot {\bf k} \right) - \frac{c^2}{\hbar \omega} \left({\bf p}_e \cdot {\bf k} \right) \, , 
\label{2.5} \end{equation}
with $\omega_e = \epsilon_e / \hbar$, and 
\begin{equation}
F (\tau, \tau') = c^2 k_C^2 \left( U^2 - \delta a^2 \right) + 2 c k_C \left(\omega_e U - \frac{c}{\hbar} {\bf p}_e \cdot {\bf a} \right) \, , 
\label{2.6} \end{equation}
The quantity $g' (\tau, \tau')$ is determined by an expression identical to (\ref{2.5}), but where $\omega$ and ${\bf k}$ are replaced by $\omega'$ and ${\bf k'}$. Here, we also have used: $\delta a^2 = a^2 - \left< a^2 \right>$. This allows us to make implicit use of the concept of electron effective mass in an electromagnetic wave, defined as $m_* = m \sqrt{1 + \left< a^2 \right>}$, which is relevant to classical and quantum processes in laser-plasma interactions \cite{marklund,christian}. In terms of this effective mass, eq. (\ref{2.3c}) simply reads $\epsilon_e^2 = p_e^2 c^2 + m_*^2 c^4$.

It should be noticed that eq. (\ref{2.4}) is valid under the assumption that the fast time-scales of the electron wavefunction are concentrated in the phase function $\theta$, defined by eq. (\ref{2.3b}). This is strictly valid for photon energies well below the electron rest energy, $\hbar \omega \ll m_* c^2$. The exact expression of eq. (\ref{2.5}) would contain second derivative terms in $\partial^2 \Phi / \partial \tau^2$, $\partial^2 \Phi / \partial \tau^{'2}$ and $\partial^2 \Phi / \partial \tau \partial \tau'$, which are neglected. We assume, due to this time-scale argument, that they will not  change the main qualitative features of the present results. For a more complete analysis of second derivative terms see \cite{eliezer,haas2}. 
We now search for split solutions of the form
\begin{equation}
\Phi (\tau, \tau') = L (\tau) T (\tau') \, , 
\label{2.7} \end{equation}
where $\tau$ and $\tau'$ are the temporal variables associated with longitudinal and transverse field oscillations. 
Separation of variables then leads to two similar equations, of the form
\begin{equation}
2 i g (\tau, \tau') \frac{d L}{d \tau} = G (\tau) L \, , \quad 2 i g' (\tau, \tau') \frac{d T}{d \tau} = G' (\tau') T \, .
\label{2.8} \end{equation}
with the new functions
\begin{equation}
G (\tau) = c^2 k_C^2 U^2 + 2 c k_C \omega_e U \, , 
\label{2.9} \end{equation}
and
\begin{equation}
G' (\tau') = - c^2 k_C^2 \delta a^2 - 2 \frac{c^2 k_C}{\hbar} \left( {\bf p}_e \cdot {\bf a} \right) \, .
\label{2.9b} \end{equation}
Notice that, by definition, the electrostatic potential $U$ only depends on $\tau$, and the electromagnetic potential ${\bf a}$ only depends on $\tau'$.
In these equations, we have used the obvious relation $F (\tau, \tau') = G (\tau) + G' (\tau')$. 
Although the separation of variables in eqs. (\ref{2.8}) is not complete, their formal solution 
is straightforward, and can be stated as
\begin{equation}
L (\tau ) = L_0 \exp \left[ - \frac{i}{2} \int_0^\tau \frac{G (z)}{g (z, \tau')} d z \right] \, ,
\label{2.10} \end{equation}
and 
\begin{equation}
T (\tau' ) = T_0 \exp \left[ - \frac{i}{2} \int_0^{\tau'} \frac{G' (z')}{g' (\tau, z')} d z' \right] \, ,
\end{equation}
where $L_0$ and $T_0$ are constants.
From the above results, we are then able to obtain the electron wavefunction solution, which takes the form
\begin{equation}
\psi ({\bf r}, t) = \psi_0 e^{i \theta} e^{i S (\tau, \tau')} \, , 
\label{2.11} \end{equation}
where $\psi_0 = L_0 T_0$, and 
\begin{equation}
S (\tau, \tau') = - \int_0^\tau \frac{G (z)}{2 g (z, \tau')} d z -  \int_0^{\tau'} \frac{G' (z')}{2 g' (\tau, z')} d z' \, .
\label{2.11b} \end{equation}

\section{Recoil formula}

These solutions can be used to describe Compton scattering processes in the nonlinear regime, as shown next. In order to understand their physical meaning, we introduce a simplifying assumption, by assuming waves with moderate amplitudes, such that we can neglect the nonlinear term in $G (\tau)$. This corresponds to $| U | \ll \omega_e / c k_C \sim 1$. We also assume an intense laser field such that the quadratic term in $\delta a^2$ in $G' (\tau')$ can eventually be dominant. Furthermore, we consider sinusoidal potential oscillations, such that $U (\tau) = U_0 \cos (\omega \tau)$ and $a (\tau') = a_0 \cos (\omega' \tau')$. Further assuming that $g (\tau, \tau')$ and  $g' (\tau, \tau')$ are or order $\omega_e$, we can reduce eq. (\ref{2.11b}) to the following simple expression
\begin{equation}
S (\tau, \tau') = \beta \sin (\omega \tau) - \beta' \sin (\omega' \tau') - \beta" \sin (2 \omega' \tau')  \, ,
\label{3.1} \end{equation}
with the three normalized frequencies 
\begin{equation}
\beta = 2 \frac{c k_C}{\omega} U_0 \, , \quad \beta' = 2 \frac{c^2 k_C}{\hbar \omega' \omega_e} {\bf p}_e \cdot {\bf a}_0 \, , \quad
\beta" = \frac{c^2 k_C^2}{2 \omega' \omega_e} a_0^2 \, .
\label{3.1b} \end{equation}
This leads to the following final expression for the electron wavefunction in the presence of two waves
\begin{equation}
\psi ({\bf r}, t) = \psi_0 e^{i \theta} \sum_{n, n',n"} J_n (\beta) J_{n'} (\beta') J_{n"} (\beta"/2) e^{ i  \theta_s (n, n', n")} \, .
\label{3.2} \end{equation}
where the phase function resulting from the existence of the two waves is determined by
\begin{equation}
\theta_s (n, n', n") = - n \omega \tau + (n'+2n") \omega' \tau' \, .
\label{3.2ab} \end{equation}
From this phase function, we can easily realise that new energy and momentum states of the electron are possible, when one of the terms contained in the sum of eq. (\ref{3.2}) becomes equal to a new electron phase function
\begin{equation}
\theta' =  - \frac{1}{\hbar} \left( \epsilon'_e t - {\bf p'}_e \cdot {\bf r} \right) \, ,
\label{3.2b} \end{equation}
where the new values of the electron energy and momentum $\epsilon'_e$ and ${\bf p'}_e$ are determined by the identity
$\theta' = \theta + \theta_s (n, n', n")$.
In explicit terms, this phase identity defines the energy and momentum conservation relations
\begin{eqnarray}
\epsilon'_e = \epsilon_e + n \hbar \omega - (n' + 2 n") \hbar \omega' \, , \nonumber \\
{\bf p'}_e = {\bf p}_e + n \hbar {\bf k} - (n' + 2 n") \hbar {\bf k'} \, .
\label{3.3b} \end{eqnarray}
Such relations set the conditions for the occurrence of Compton scattering processes involving both plasmons (with frequency $\omega$) and photons (with frequency $\omega'$). When the initial and final electron energy states are nearly identical $\epsilon_e \sim \epsilon'_e$, and the plasma is strongly underdense, $\omega' \gg \omega_p$, this corresponds to the decay of one photon (for $n' + 2 n" = 1$) into a large number of plasmons $n \gg 1$, such that $n \simeq \omega' / \omega_p$. Conversely, we can eventually convert $n \gg 1$ plasmons into a single high frequency photon. Higher order processes, involving more than one photon, are also possible. 

From this analysis, we can also retrieve the recoil effect suffered by an electron upon scattering. For this purpose, we assume an electron nearly at rest, with initial momentum, $p_e \ll m_* c$, such that it can be neglect. In this case, we use $\epsilon_e \simeq m_* c^2$ in the first eq. (\ref{3.3b}),  with the effective mass $m_* = m_e \gamma_0$ and $\gamma_0 = (1 + a_0^2)^{1/2}$. In the expression for the relativistic factor $\gamma_0$, we have neglected the contributions from the electrostatic wave, which can easily be included when justified. Using $\epsilon'_e = \epsilon_e + \hbar \omega' - n \hbar \omega$, we obtain
\begin{equation}
p^{'2}_e c^2 = \left[ m_a c^2 + \hbar (\omega' - n \omega) \right]^2 - m_a^2 c^4 \, .
\label{3.5} \end{equation}
This allows us to calculate the electron recoil momentum using the second eq. (\ref{3.3b}), ${\bf p'}_e = n \hbar {\bf k} - \hbar {\bf k'}$. We get 
\begin{equation}
p^{'2}_e c^2 = c^2 \left[ n^2 \hbar^2 k^2  + \hbar^2 k^{'2} - 2 n \hbar^2 k k' \cos \varphi \right] \, ,
\label{3.5b} \end{equation}
where $\varphi$ is the angle between the momenta, ${\bf k}$ and ${\bf k'}$, of the emitted and absorbed quanta of radiation.  

We now use the photon and plasmon dispersion relations for a relativistic plasma (see, for instance, \cite{pop2011}), assuming that $v_{the}^2 / c^2 \ll 1$, where $v_{the} = \sqrt{3 T_e / m_*}$ for a thermal energy $T_e$. Equating these two expressions, we obtain
\begin{eqnarray}
\frac{1}{2 \omega'} \left[ n \omega \left(1 - \frac{c^2}{v_{the}^2} \right) + \frac{\omega_p^2}{2 n  \omega} \left(1 + \frac{n^2 c^2}{v_{the}^2} \right) \right] + \nonumber \\
\gamma_0 k_C \left( \frac{c}{n \omega} - \frac{c}{\omega'} \right) = 1 - \alpha(\omega) \cos \varphi \, ,
\label{3.6} \end{eqnarray}
with the auxiliary function 
\begin{equation} 
\alpha (\omega) = \frac{1}{\omega \omega'} \frac{c}{v_{the}} \sqrt{(\omega^2 - \omega_p^2/\gamma_0)(\omega^{'2} - \omega_p^2/\gamma_0)}  \, .
\label{3.6b} \end{equation}
This expression gives the electron recoil under Compton plasmon scattering. It corresponds to the absorption of one photon with frequency $\omega$, and emission of $n$ plasmons with frequency $\omega'$. A more familiar expression can be obtained if we neglect the plasma dispersion effects associated with the two terms under square brackets in (\ref{3.6}) and take the limit of $\alpha (\omega) \rightarrow 1$. We then would get 
\begin{equation}
\frac{c}{n \omega} - \frac{c}{\omega'} = \frac{1}{k_C} \frac{(1 - \cos \varphi)}{\sqrt{1 + a_0^2}} \, .
\label{3.7} \end{equation}
This particular case exactly mimics the well-known formula for nonlinear Compton scattering of photons in vacuum. 
An illustration of this recoil formula is shown in Figure 1. But, in general, eq. (\ref{3.6b}) should be used.

\begin{figure}
\includegraphics[angle=0,scale=0.3]{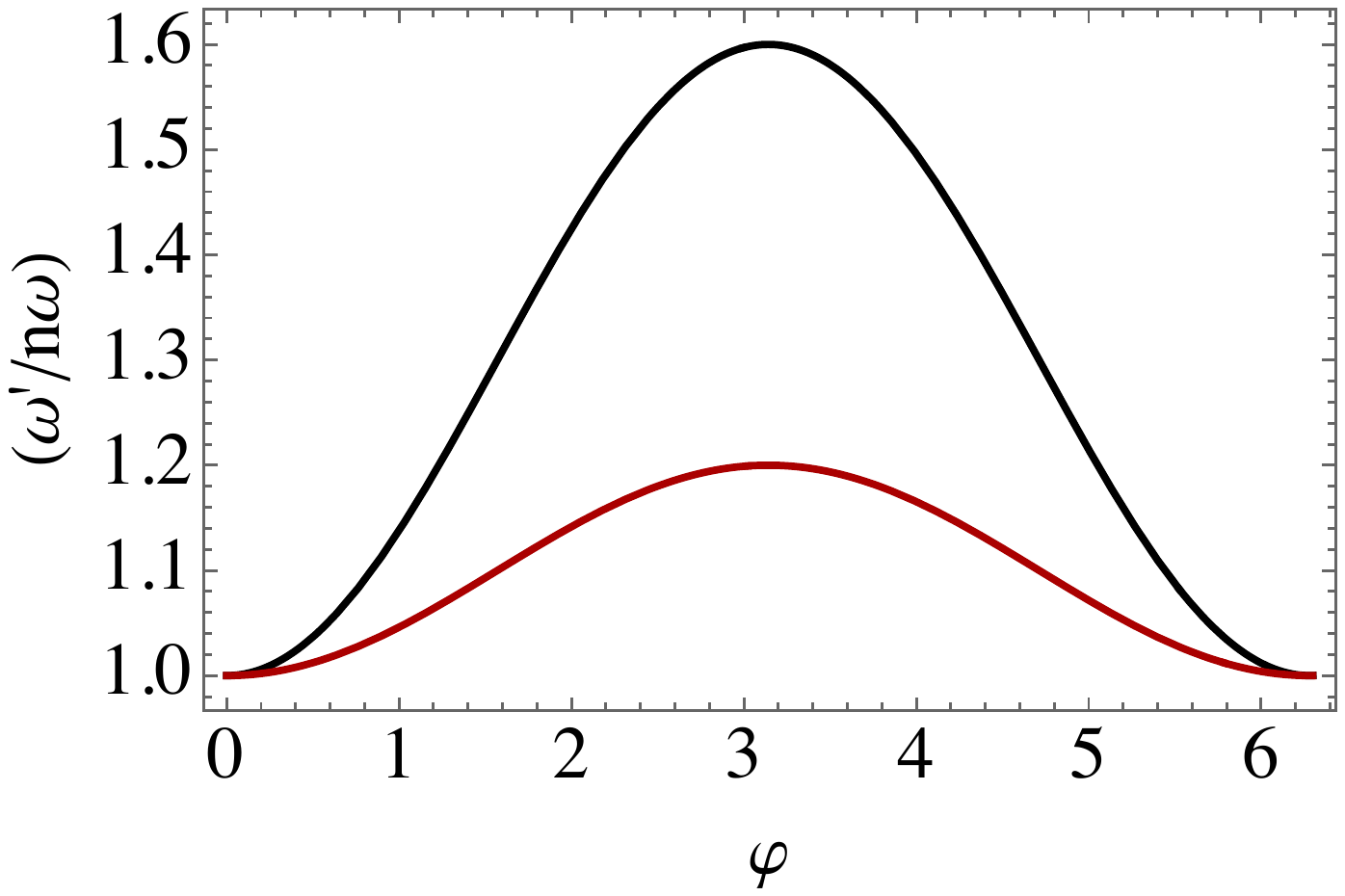}
\caption{\label{fig1}  {\sl Frequency of $n \gg 1$ scattered plasmons, of a photon with frequency $\omega' = c k_C$, for $a_0 = 3$ (black curve) and $a_0 = 10$ (red curve), after nonlinear Compton scattering. }}
\end{figure}

\section{Scattering Probability}

Let us now discuss the probability for these Compton scattering processes to occur, with energy conversion between electrostatic and electromagnetic waves. This is dictated by the amplitude $P_{n,n',n"} \equiv \left| \psi_{n,n',n"} \right|^2$, satisfying the conservation relations (\ref{3.3b}), and defined as
\begin{equation}
P_{n,n',n"} = \left| J_n (\beta) \right|^2 \left| J_{n'} (\beta') \right|^2 \left| J_{n"} (\beta") \right|^2 \, .
\label{3.4} \end{equation}
This probability allows us to define the scattering cross-section $\sigma$, as $d \sigma/ d \Omega = P_{n,n',n"} / F_0$, where $d \Omega$ is the element of solid angle, $F_0 = I_0 / \hbar \omega_0$ is the incident photon flux, and $I_0 \propto a_0^2$ is the radiation intensity. It is useful to derive a more explicit expression for the probability, valid when the plasma wave amplitude is very small, $U_0 \ll 1$. Using the asymptotic expansion of the Bessel functions for small arguments, assuming that $a_0^2 \gg 1$ and  $n' = 0$, $n" = 1$, we can then approximately write
\begin{equation}
P_{n,1} = \left( \frac{1}{2^{n-1}} \frac{c k_C}{n! \, \omega} U_0 \right)^{2n}  \left| J_1(\beta") \right|^2 \, .
\label{3.4c} \end{equation}
where $\beta" \propto a_0^2$, as defined in eq. (\ref{3.1b}). In this expression, we should use the value of $U_0$ associated with a state of $n$ plasmons at the frequency $\omega \simeq \omega_p$. Using the expression for the energy of an electron plasma wave with this amplitude, and equating it to $\hbar n \omega$, we get
\begin{equation}
U_0 = \sqrt{ \frac{\hbar n \omega \omega_p}{2 \epsilon_0 k^2}} \sim v_{the} \sqrt{\frac{\hbar n}{2 \epsilon_0}} \, .
\label{3.4b} \end{equation}
This completely determines the probability for the Compton scattering of plasmons, in an underdense plasma, when no plasmons are present except those resulting directly from the elementary Compton process. This transition probability is illustrated in Figure 2, for two different values of the number of scattered plasmons $n$. We notice that the probability strongly decreases and eventually becomes negligible for a large number of plasmons $n \gg 1$, but it takes significant values when this number is small. 
The observed decrease of the probability when the laser intensity increases is somewhat surprising, but it results from the structure of the Volkov solutions, and is an ultimately consequence of the increase of the effective electron mass. The first maximum, observed in the figure near $a_0 = 1.4$, is shifted towards larger values with the photon frequency $\omega'$ decreases. 

On the other hand, similarly to Compton scattering of photons in vacuum, the difference between absorbed and emitted energy quanta of radiation, $\hbar (\omega' - n \omega)$ is limited by the kinetic energy of the electrons. For a thermal plasma, this difference is typically of the order to the electron thermal energy, $T_e$. However, if scattering is due to supra-thermal electrons (created for instance by the laser pulse itself), then we can eventually get $\omega' \gg n \omega$, or the inverse process of $n \omega \ll \omega'$, with a small number of $n = 1$ or $2$. This provides a broad range of physical scenarios where (direct and inverse) Compton scattering of plasmons can eventually occur with non-negligible probability.
 
\begin{figure}
\includegraphics[angle=0,scale=0.3]{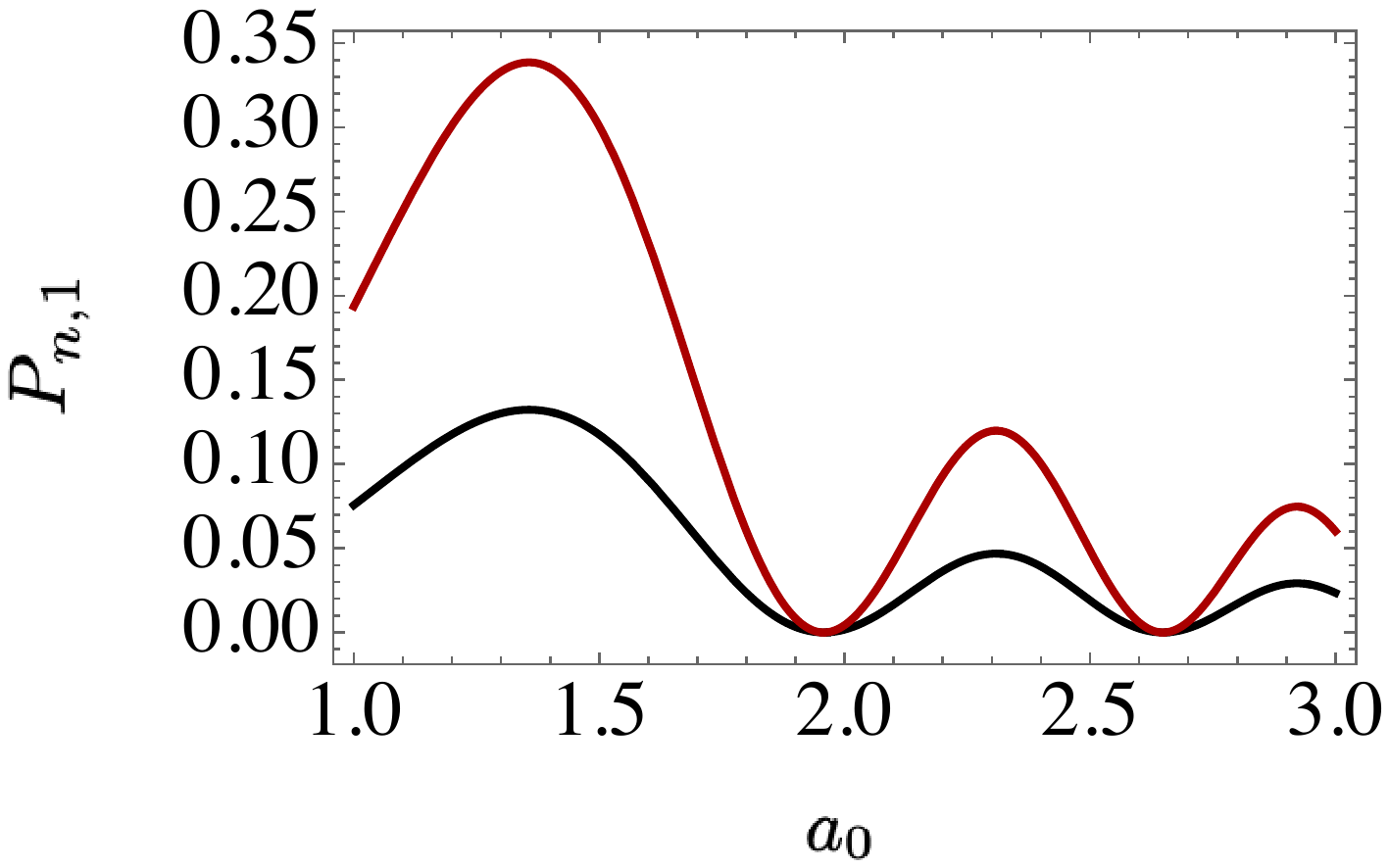}
\caption{\label{fig2}  {\sl Transition probability $a_0^2 P_{n,1}$ for plasmon scattering, when the number of plasmons is equal to $n = 1$ (red curve) and $n = 2$ (black curve, multiplied by $10^2$), as a function of the laser field amplitude $a_0$. We have assumed $U_0 = (\omega/ c k_C) \ll 1$.}}
\end{figure}

Finally, replacing eq. (\ref{3.4b}) in (\ref{3.4c}), we can easily demonstrate that the scattering probability is proportional to the n-th power of the electron plasma density $n_0$, and of the classical electron radius $r_e = \alpha \hbar /\sqrt{2} m c$, where $\alpha$ is the fine structure constant, according to $P_{n, 1} \propto (n_0 r_e^3)^n \left| J_1(\beta") \right|^2$. Obviously, this applies to the photon-plasmon scattering described here. A similar formulation would be possible for plasmon-plasmon scattering, and would then lead to a probability $P_{1,1} \propto n_0^{2/3} r_e^2$. For each particle in the plasma, this probability would be of order $r_e^2$, in analogy with the usual Compton scattering of photons in vacuum.

\section{Conclusions}

In conclusion, we have extended the concept of Compton scattering to the case of plasmons. Our description was based on the Volkov solutions for electrons in the field of two waves, an electromagnetic and an electrostatic wave in plasmas. We were able to derive the electron recoil formula for this new Compton process, as well as the expressions for the probability to produce $n$ plasmons upon scattering. It should be noticed that, when the photon frequency is large with respect to the plasma frequency, and the number $n$ increases, the scattering probability significantly drops. Conversely, if a large amplitude electron plasma wave already exists in the medium, the inverse plasmon Compton scattering will eventually lead to the emission of high frequency photons. This could be significant for experiments on intense laser-plasma interactions. 

Spin is not present in the famous Klein-Nishina formula for the original Compton scattering of photons, and is also ignored in the this work. It can nevertheless become important in the presence of intense laser beams. For spin-dependent Compton scattering see \cite{ahrens} and references therein. 

The simple quantum model considered here can eventually be extended to a more complete quantum kinetic description of electron plasma waves \cite{brodin}.  In a recent work, this kinetic approach revealed the existence of multi-plasmons resonances associated with electron Landau damping \cite{brodin2}. In contrast here, the multi-plasmon resonaces also involve the presence of electromagnetic waves, which are absent in this work. In  our case, electrons interact resonantly with photons and plasmons, and not just with plasmons. Moreover, our model concerns single-particle processes and, for that reason is completely distinct from stimulated scattering processes. A quantum kinetic theory could in principle be able to include both single-particle and stimulated scattering processes in a coherent description, to be considered in the future.

In our view, the proposed concept of Compton scattering of plasmons provides a natural extension of the celebrated Compton photon scattering, and could contribute to a better understanding of the radiation processes in plasma physics.

\begin{acknowledgements}

F.H. would like to thank the Instituto Superior Tecnico, and in particular, the Group of Lasers and Plasmas (GoLP), for hospitality and support during his short visit to Lisbon. 
He also would like to acknowledge financial support of CNPq (Conselho Nacional de Desenvolvimento Cient\'ifico e Tecnol\'ogico), Brazil. J.T.M. would like to thank Jorge Vieira and Hugo Ter\c cas for stimulating discussions on laser-plasma interactions and quantum plasma processes, respectively.

\end{acknowledgements}


\begin{thebibliography}{99}

\bibitem{compton}
A. H. Compton, ``A Quantum Theory of the Scattering of X-rays by Light Elements", {\sl Phys. Rev.} {\bf 21}, 483 (1923).

\bibitem{heitler}
W. Heitler, {\sl The Quantum Theory of Radiation}, Oxford University Press, London, 3rd. edition (1954)

\bibitem{itzykson}
C. Itzykson and J.-B. Zuber, {\sl Quantum Field Theory}, McGraw-Hill, New York (1980).

\bibitem{mbook}
J. T. Mendon\c ca, {\sl The Quantum Nature of Light}, Institute of Physics Publishing, Bristol (2022).

\bibitem{blumenthal}
G. R. Blumental and R.  J. Gould, ``Bremsstrahlung, Synchrotron Radiation, and Compton Scattering of High-Energy Electrons Traversing Dilute Gases", 
{\sl Rev. Mod. Phys.} {\bf 42}, 237 (1970).

\bibitem{zhang}
Y. Zhang, J.-J. Geng and Y.-F. Huang, ``Inverse Compton Scattering Spectra of Gamma-Ray Burst Prompt Emission", {\sl AstroPhys. J.} {\bf 877}, 89 (2019).

\bibitem{hartemann}
F. V. Hartemann, A. L.  Troha, H. A. Baldis, A.  Gupta, A. K. Kerman, E. C. Landahl, N. C.  Neville Jr. and J. R. Van Meter,  ``High-Intensity Scattering Processes of Relativistic Electrons in Vacuum and their Relevance to High-Energy Astrophysics", 
{\sl AstroPhys. J. Suppl. Series} {\bf 127}, 347 (2000).

\bibitem{brown}
L. S. Brown and T. W. B. Kibble, ``Interaction of Intense Laser Beams with Electrons", {\sl Phys. Rev.} {\bf 133}, A705 (1964).

\bibitem{nikishov}
A. I. Nikishov and V. I. Ritus, ``Quantum Processes in the Field of a Plane Electromagnetic Wave and in a Constant Field", 
{\sl Sov. Phys. JETP} {\bf 19}, 529 (1964).

\bibitem{eberly}
Z. Fried and J. H. Eberly, ``Scattering of a High-Intensity, Low-Frequency Electromagnetic Wave by an Unbound Electron", 
{\sl Phys. Rev.} {\bf 136}, B871 (1964)

\bibitem{dipiazza}
F. Mackenroth and A. Di Piazza, ``Nonlinear Compton Scattering in Ultrashort Laser Pulse", 
{\sl Phys. Rev. A} {\bf 83}, 032106 (2011).

\bibitem{seipt}
D. Seipt, V. Kharin, S. Rykovanov, A. Surzhykov and S. Fritzsche,``Analytical Results for Nonlinear Compton Scattering in Short Laser Pulses", 
{\sl J. Plasma Phys.} {\bf 82}, 655820203 (2016).

\bibitem{thoma}
F. Del Gaudio, R. A. Fonseca, L. O. Silva, and T. Grismayer, ``Plasma Wakes Driven by Photon Bursts via Compton Scattering",
{\sl Phys. Rev. Lett.} {\bf 125}, 265001 (2020).

\bibitem{corels}
J. W. Yoon, Y. G. Kim, I. W. Choi, J. H. Sung, H. W. Lee, S. K. Lee and C. H. Nam,
``Realization of Laser Intensity over $10^{23} \, W / cm^2$",
 {\sl Optica} {\bf 8}, 630 (2021).

\bibitem{volkov}
D. M. Wolkow, ``\"Uber eine Klasse von L\"osungen der Diracschen Gleichung", 
{\sl Zeit. Physik} {\bf 94}, 250 (1935).

\bibitem{pre2011}
J. T. Mendon\c ca and A. Serbeto, ``Volkov Solutions for Relativistic Quantum Plasmas", 
{\sl Phys. Rev. E} {\bf 83}, 026406 (2011).

\bibitem{eliezer}
E. Raicher and S. Eliezer, ``Analytical Solutions of the Dirac and Klein-Gordon Equations  in Plasma Induced by High-Intensity Laser",
{\sl Phys. Rev. A} {\bf 88}, 022113 (2013).

\bibitem{haas}
F. Haas and M. A. A. Manrique, ``Effective Photon Mass and Exact Translating Quantum Relativistic Structures", 
{\sl Phys. Plasmas} {\bf 23}, 042102 (2016).

\bibitem{marklund}
C. Harvey, T. Heinzl, A. Ilderton and M. Marklund, ``Intensity-dependent Mass Shift in a Laser Field: Existence, Universality and Detection", 
{\sl Phys. Rev. Lett.} {\bf 109}, 100402 (2012).

\bibitem{christian}
C. Kohlf\"urst, H. Gies and R. Alkofer, ``Effective Mass Signatures in Multiphoton Pair Production",
{\sl Phys. Rev. Lett.} {\bf 112}, 050402 (2014).

\bibitem{haas2}
F. Haas, J. T. Mendon\c ca and H. Ter\c cas, ``Quantum Landau Damping in the Nonlinear Regime", in preparation (2023).

\bibitem{pop2011}
J. T. Mendon\c ca, ``Wave Kinetics in Relativistic Quantum Plasmas", {\sl Phys. Plasmas} {\bf 18}, 062101 (2011).

\bibitem{ahrens}
S. Ahrents and C-P. Sun, ``Spin in Compton Scattering with Pronounced Polarization dynamics", {\sl Phys. Rev. A} {\bf 96}, 063407 (2017).

\bibitem{brodin}
G. Brodin and J. Zamanian, ``Quantum Kinetic Theory of Plasmas", {\sl Rev. Mod. Plasma Phys.} {\bf 6}, 4 (2022).

\bibitem{brodin2}
A. P. Misra and G.  Brodin, ``Wave-Particle Interactions in Quantum Plasmas",
{\sl Rev. Mod. Plasma Phys.} {\bf 6}, 5 (2022).

\end{thebibliography}
\end{document}